XPS analysis of molecular contamination and sp2 amorphous carbon on oxidized (100) diamond

Ricardo Vidrio[1,ǂ,*], Daniel Vincent[2,ǂ], Benjamin Bachman[3], Cesar Saucedo[3], Maryam Zahedian[2], Zihong Xu[4], Junyu Lai[5], Timothy A. Grotjohn[6], Shimon Kolkowitz[7,8], Jung-Hun Seo[5], Robert J. Hamers[3], Keith G. Ray[9], Zhenqiang Ma[2], Jennifer T. Choy[1,2,*]

[1]1500 Engineering Dr, Madison WI 53706, Department of Nuclear Engineering and Engineering Physics, University of Wisconsin-Madison

[2]1415 Engineering Dr, Madison WI 53706, Department of Electrical and Computer Engineering, University of Wisconsin-Madison

[3]1101 University Ave, Madison WI 53706, Department of Chemistry, University of Wisconsin-Madison

[4]480 Lincoln Dr, Madison WI 53706, Department of Mathematics, University of Wisconsin-Madison

[5]135 Bell Hall, Buffalo NY 14260, Department of Materials Design and Innovation, University at Buffalo

[6]474 S Shaw Ln, East Lansing MI 48824, Electrical and Computer Engineering, Michigan State University

[7]1150 University Ave, Madison WI 53706, Department of Physics, University of Wisconsin-Madison

[8]366 Physics North MC 7300, Berkeley CA 94720, Department of Physics, University of California-Berkeley

[9]7000 East Ave, Livermore CA 94550, Lawrence Livermore National Laboratory

[ǂ]These authors contributed equally to this work.

## Abstract

The efficacy of oxygen (O) surface terminations on diamond is an important factor for the performance and stability for diamond-based quantum sensors and electronics. Given the wide breadth of O-termination techniques, it can be difficult to discern which method would yield the highest and most consistent O coverage. Furthermore, the interpretation of surface characterization techniques is complicated by surface morphology and purity, which if not accounted for will yield inconsistent determination of the oxygen coverage. We present a comprehensive approach to consistently prepare and analyze oxygen termination of surfaces on (100) single-crystalline diamond. We report on X-ray Photoelectron Spectroscopy (XPS) characterization of diamond surfaces treated with six oxidation methods that include various wet chemical oxidation techniques, photochemical oxidation with UV illumination, and steam oxidation using atomic layer deposition (ALD). Our analysis entails a rigorous XPS peak-fitting procedure for measuring the functionalization of O-terminated diamond. The findings herein have provided molecular-level insights on oxidized surfaces in (100) diamond, including the demonstration of clear correlation between the measured oxygen atomic percentage and the presence of molecular contaminants containing nitrogen, silicon, and sulfur. We also provide a comparison of the $sp^2$ carbon content with the O1s atomic percentage and discern a correlation with the diamond samples treated with dry oxidation which eventually tapers off at a max O1s atomic percentage value of 7.09 ± 0.40%.

*Corresponding authors: jennifer.choy@wisc.edu (Jennifer Choy)
vidrio@wisc.edu (Ricardo Vidrio)

Given these results, we conclude that the dry oxidation methods yield some of the highest oxygen amounts, with the ALD water vapor technique proving to be the cleanest technique out of all the oxidation methods explored in this work.

## 1. Introduction

Diamond is a material that is renowned for its unique properties such as extreme hardness [1] [2], high thermal conductivity [3], chemical inertness [4], and biocompatibility [5]. In particular, the study of surface functionalizations on single-crystalline diamond (SCD) for the purposes of quantum sensing with color centers is an active area of research today. For example, Xie et. al have engineered a process where SCD sample surfaces were modified to immobilize biological sensing targets which resulted in optimal coherence times for NV sensing [6]. However, this is also a field that extends to other color centers, as oxygen (O) terminations have been shown to be favorable towards maintaining a negative charge state in silicon vacancy (SiV) centers [7].

Although different types of surface terminations have been studied on diamond, including nitrogen [8], hydrogen [9], fluorine [10], and even silicon [11], oxygen (O) terminations on diamond remains as some of the most widely adopted. The study of O-terminations on SCD samples prove pertinent in quantum sensing applications, as O-terminations on shallow NV centers have been shown to improve spin coherence times [12] [13], which in turn results in magnetometers with improved sensitivity.

In this work, we describe and demonstrate experimental approaches to consistently prepare and interpret the results of oxidizing SCD surfaces cut in the (100) configuration. We apply wet chemistry and dry oxidation methods to O terminate the diamond surface and use XPS to interpret the surface functionalizations. We describe in detail our peak-fitting procedure for quantifying the surface chemistry of O-terminated and hydrogen (H) terminated samples and emphasize that the reporting of residuals along with certain peak parameters (including peak energies, line shapes, and full-width-half-maximum (FWHM) values) is imperative for evaluating the quality of the fit and comparison across different data sets. One of our findings indicate that the presence of molecular contaminants on the diamond surface resulted in an exaggerated O atomic percentage.

We group diamond oxidation methods into two different categories: wet or dry chemistry oxidation. As the name implies, wet chemistry oxidation (WCO) involves the use of acid solutions to induce O-termination on the diamond surface, while dry oxidation (DO) methods achieve O-termination with gases. By far, the most prevalent of the oxidation methods reported has been exposure of diamond surfaces to aqueous mineral acids at elevated temperatures. As shown by Damle et. al [14], Wang et. al [15], and Maier et. al [16], a mixture of $H_2SO_4$ and $HNO_3$ acids are a commonly used and generally effective means of diamond surface oxidation for (100) SCD. To our knowledge, there is no consensus on the mechanism responsible for diamond oxidation using $H_2SO_4$ and $HNO_3$ acid mixtures, but Li et. al identified $NO_2^+$ as the likely oxidizing agent based on their work with polycrystalline and diamond powders in $H_2SO_4$:$HNO_3$ acid baths [17].

*Table 1: Compilation of experimental oxidation methods and results that have been used on (100) single-crystalline diamond.*

| Category of treatment | Oxidation Method | O1s At % | Reference |
|---|---|---|---|
| Wet | $H_2SO_4$:$HNO_3$ (3:1) for 4 hours at 140˚C | 13.73 ± 5.42 | Damle et. al (2020) [14] |
| Wet | Hydrogen-termination, $H_2SO_4$:HNO3 bath, NaOH bath, and HCl bath | 16 | Wang et. al (2011) [15] |
| Wet | Tri-acid clean/$HClO_4$:$H_2SO_4$:$HNO_3$ (1:1:1) for 2 hours | 8.5 | Cui et. al (2013) [59] |
| Wet | Tri-acid Clean/ $HClO_4$:$H_2SO_4$:$HNO_3$ (1:1:1) for 1 hour | 4.82 ± 0.74 | Sangtawesin et. al (2019) [12] |
| Wet | Tri-acid clean/ $HClO_4$:$H_2SO_4$:$HNO_3$ (1:3:4) for 2 hours | 4.94 | Alba et. al (2020) [18] |
| Wet | H-termination, Sulfo-Chromic Acid at 230-250˚C | 6.8 | Klauser et. al (2010) [19] |
| Dry | H-termination, O plasma | 6.64 | Klauser et. al (2010) [19] |
| Dry | H-termination, Thermal Oxidation at atmospheric conditions at 700 ˚C for 5 minutes | 5.85 | Klauser et. al (2010) [19] |
| Dry | UV-Ozone illumination with atmospheric conditions | 5.27 | Klauser et. al (2010) [19] |
| Wet and Dry | Tri-acid Clean/ $HClO_4$:$H_2SO_4$:$HNO_3$ (1:1:1) for 1 hour + oxygen anneal + piranha clean | 6.17 ± 0.99 | Sangtawesin et. al (2019) [12] |

A common acid treatment used in the diamond community also includes perchloric acid ($HClO_4$) in a 1:1:1 volumetric ratio with $H_2SO_4$ and $HNO_3$ acids [12] [18]. In this tri-acid mixture, the chemical reaction proceeds according to

$$2H_2SO_4 + HNO_3 + HClO_4 \leftrightarrow NO_2^+ + H_2O + HSO_4^- + ClO_4^- + H_3SO_4^+ \quad (1)$$

When compared to $H_2SO_4$:$HNO_3$, the tri-acid mixture now includes the presence of two oxidizing agents, $NO_2^+$ and $H_3SO_4^+$, both of which could participate in the diamond oxidation process.

Although the nitric-sulfuric mixtures are prolific in their use of diamond oxidation, there are other acids which have also been employed towards O-terminating diamond. These include sulfo-chromic acid [19], aqua regia [20], and $CrO_3$:$H_2SO_4$ [21]. The results of subjecting the diamond surface to these acids are shown in Table 1. An alternative to liquid acids for oxidizing diamond surfaces includes exposure to O gases, often under photochemical (via UV exposure) or plasma discharge conditions, as shown in [19] and [22] in Table 1. The efficacy of this method relies on the direct adsorption of O on the diamond surface, as was demonstrated by Enriquez et. al [23].

The most common approach to quantify the extent of the O coverage on the SCD surface is to use XPS to determine the total percentage of O atoms, relative to C atoms, that are present on the surface. Table 1 compiles the experimental methods and XPS results for representative prior work on oxidation of (100) SCD samples. Although other work exists that quantifies oxygen content in other ways besides oxygen atomic percentage, such as in MegaLangmuir (ML) [24], we chose to restrict our comparisons to literature with XPS data.

While the focus of this work is on (100) SCD, we note that the oxidation techniques explored in this work and elsewhere are applicable to other crystal orientations, including (111) SCD, which is conducive to improved contrast in optically detected magnetic measurements for NV centers [25] [26]. In the extensive study by Klauser et. al, whose oxidation techniques are shown in Table 1, wet and dry oxidation methods were performed on chemical vapor deposition (CVD) grown and natural type IIb diamond of (100) and (111) orientation. It was found that with identical oxidation methods that both (100) and (111) had similar O1s atomic percentages, within 0.5%. The main difference between both of these orientations lies in the presence of the functionalization groups, as the O plasma treatment for the natural type IIb (100) diamonds resulted in 55% more singly bonded O groups when compared to the type IIb (111) diamonds [19].

The inconsistencies in the reporting of XPS outcomes in literature reflect the dependence of oxidation on a multitude of factors as well as a lack of a set of established protocols that can consistently prepare and analyze diamond surfaces using XPS. In this experimental study, we apply a range of wet and dry surface treatment methods on (100)-SCDs and analyze their surface using XPS and atomic force microscopy (AFM). Our approach enabled us to identify key chemical parameters, namely $sp^2$ C content and the presence of molecular contaminants, as critical to interpreting oxygen atomic percentage values and can reasonably account for the previously unexplained inconsistencies in literature.

## 2. Experimental Methods

### *2.1 Materials*

All diamonds used for this study are type IIa (100) SCD grown using chemical vapor deposition (CVD) by Element 6. The type of diamonds employed here fall into two categories based on their surface roughness values as measured by AFM (Figure 1). As-received SCD samples from the vendor are typically polished to Ra < 30 nm as specified by the manufacturer (Figure 1a). We refer to these samples as “regular-polished” diamonds. All regular-polished diamonds were taken as-received from the manufacturer and then had all their subsequent oxidation treatments applied. Diamonds that are referred to as “super-polished” have been mechanically polished to Ra < 1 nm (Figure 1b) and underwent an Inductively Coupled Plasma – Reactive Ion Etching (ICP-RIE) process to remove any mechanical defects due to polishing, following the procedure in [12]. The ICP-RIE consisted of 400 W of ICP power, 250 W substrate bias RF power, 25 sccm Ar, 40 sccm $Cl_2$, 8 mTorr for 30 minutes which was then followed by 700 W ICP, 100 W substrate bias, 30 sccm $O_2$, 10 mTorr for 25 minutes [9]. We estimate the etch depth to be roughly 3 microns. The regular-polished diamonds exhibit polishing marks from the manufacturers polishing, whereas the super-polished diamonds have a lack of these polishing marks due to the ICP-RIE. After the ICP-RIE, the super-polished samples were then tri-acid cleaned and underwent XPS survey analysis to ensure a clean surface. Afterwards, the samples were vacuum annealed at 800°C with an annealing recipe taken from [12], subsequently tri-acid cleaned and then had their respective oxidation technique applied to them.

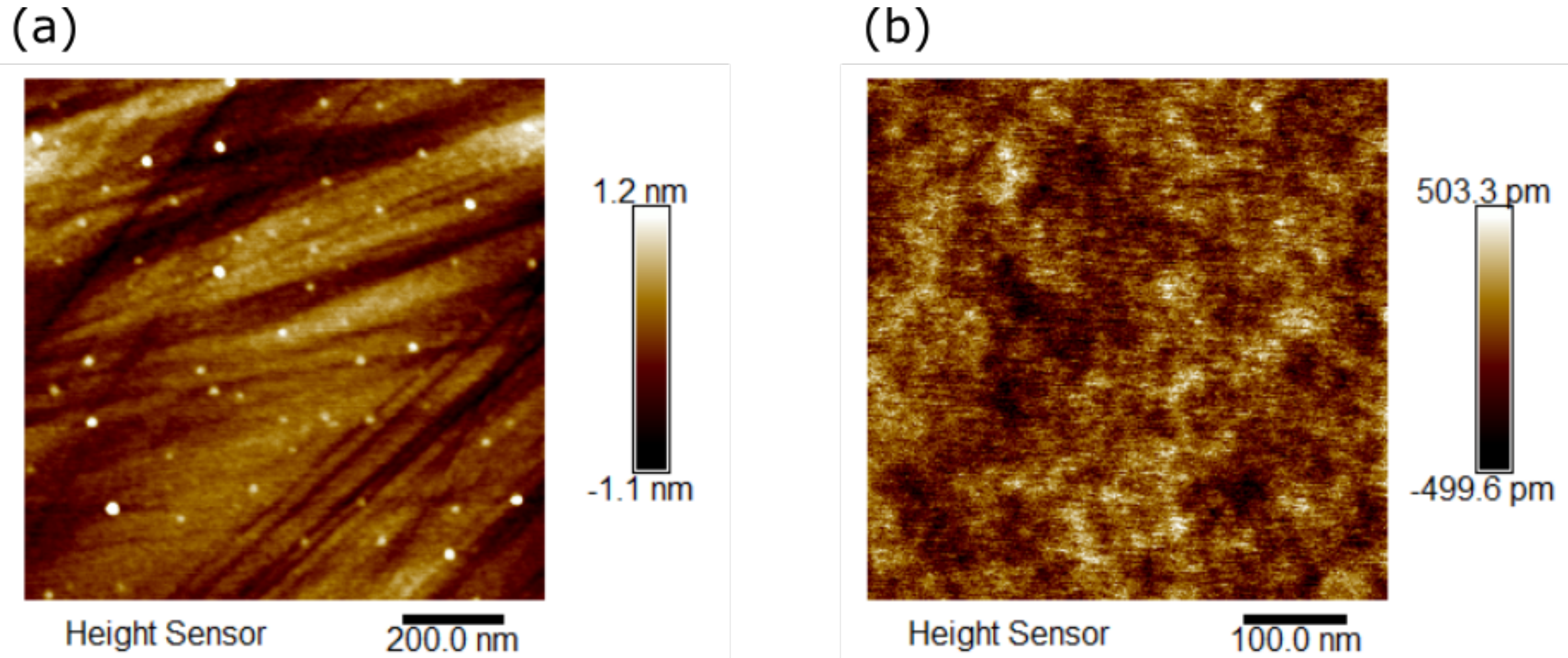

*Figure 1: AFM results of tri-acid cleaned (a) regular-polished (100) SCD surface and (b) super-polished (100) SCD surface.*

*2.2 Wet Chemistry Oxidation techniques*

All as-received diamonds were first tri-acid cleaned to remove superficial forms of $sp^2$ from the CVD growth process or any forms of adventitious carbon. We investigated three wet oxidation approaches for this study: a tri-acid clean, an $H_2SO_4$:$HNO_3$ bath, and a piranha clean. The tri-acid clean utilizes a Graham condenser and round bottom flask heater. The Graham condenser is required as a safety precaution to protect against the vapors of the boiling acid mixture, which holds steady at 450˚C. Three acids, $HClO_4$, $H_2SO_4$, and $HNO_3$ are mixed in a 1:1:1 volumetric ratio. The acid mixture is then put in a 100 mL round bottom flask and the diamond sample is placed in the solution. The round bottom flask is then inserted to the Graham condenser and the mixture is allowed to sit and boil for one hour. The piranha clean procedure consists of first tri-acid cleaning the as-received diamonds to remove most organic contamination and residual $sp^2$ carbon from the CVD growth process. Following the tri-acid clean, the diamonds are then submerged in a 3:1 $H_2SO_4$:$H_2O_2$ piranha clean mixture for five hours at 110˚C [27]. Similarly, the $H_2SO_4$:$HNO_3$ clean starts off by subjecting the as-received diamonds to a tri-acid clean. Afterwards, diamond samples were placed in a $H_2SO_4$:$HNO_3$ bath in a 9:1 volumetric mixture at 90˚C for 9 hours [28].

*2.3 H-termination and subsequent treatments*

Before H-termination took place, as-received diamond samples were first tri-acid cleaned to remove any contaminants and excess $sp^2$ C from the CVD process. SCD samples were H-terminated by placing the diamonds within an $H_2$ plasma chamber for roughly 15 minutes exposure. We then replicated the procedure by Wang et. al [15] to subject H-terminated diamond samples to first a $H_2SO_4$:$HNO_3$ bath at 90˚C, followed by a 0.1M bath of NaOH at 90˚C for 2 hours, and then submerged in a 0.1M HCl bath at 90˚C for 2 hours. We refer to this method as the H-terminated plus carboxylation (HPC) method.

*2.4 ALD and UV/Ozone treatments*

The atomic layer deposition (ALD) treatment utilized a Savannah S200 Series ALD System from Ultratech/Cambridge Nanotech. The process used a high-purity $H_2O$ precursor to treat the surface over 15 (or 30 cycles) at a temperature of 250℃. Each cycle consisted of a $H_2O$ pulse comprised of 0.015 seconds followed by a 4-second hold before purging the chamber. The UV/Ozone treatment was done using a Samco Model UV-1 UV Ozone Cleaner. The samples were placed on a substrate heater at 200˚C, while flowing 0.75L.min of oxygen, corresponding to a concentration of roughly 11 g/$m^3$ of ozone, for 35 minutes.

*2.5 Surface Characterization Techniques*

Diamond samples underwent XPS analysis with a Thermo Scientific K-Alpha X-Ray Photoelectron Spectrometer. XPS protocol relied on collecting three different points on the surface of each sample with a 200-µm spot size. For each set of survey spot measurements, one C1s and O1s spectra was measured and analyzed, corresponding to the first spot on the SCD surface. A 20-eV pass energy was chosen for all narrow scans, alongside a 0.1 eV step size. Peak fitting was performed on the C1s XPS data using CasaXPS analysis software [29]. To compensate for surface charging effects during XPS data-acquisition, the flood gun was always utilized throughout the entirety of the measurement. To analyze the $sp^2$ C, $sp^3$ C, and oxygen functionalizations on the diamond surface, we exclusively worked in peak-fitting the narrow scan C1s spectra for all diamond samples. Although narrow O1s scan data was collected, no peak-fitting was attempted for this data set, as adventitious sources of moisture could potentially shroud the presence of the functionalizations on the diamond.

A Bruker Icon Atomic Force Microscope was used to characterize the surface roughness and topology of diamond samples. AFM was employed to ascertain the difference between regular-polished and super-polished diamond samples.

## 3. Results and discussion

### *3.1 Impact of molecular contaminants on measured O concentration on the surface*

Figure 2a shows the results of representative XPS survey spectra for a SCD surface after being subjected to a WCO. Visible peaks include the C1s and O1s regions, as well as their corresponding Auger peaks, labeled as CKLL and OKLL. Shown in the subset plot are some of the most common types of contaminants observed in the spectra which include the Si2s, Si2p, S2s, and S2p peaks. Not shown in the subplot is the N1s peak, which was also found during some measurements. From these XPS survey data results, the atomic percentages for all elements identified in the survey were then calculated via CasaXPS. Figure 2a includes data defined as "*Contaminated"* compared against "*Contaminant-free Tri-Acid Clean"*. "Contaminated" data refers to diamond samples that were found to have noticeable amounts of molecular contaminants, whereas the "Contaminant-free" spectra were found to be free of any source of these impurities. As will be discussed later, contaminated samples were processed with sub-optimal methods of sample preparation and storage, including the use of non-optima grade acids and cleanroom class grade containers.

Of particular concern is how the O1s atomic percentage has a strong correlation with the amount of contaminants present on the sample. Figure 2b shows the relation between the sum of 2s and 2p atomic percentages for both silicon and sulfur, as a function of O1s atomic percentage for various initial attempts at WCO treatments. The amounts of the total silicon and sulfur contributions yield positive linear correlations with respect to the O1s atomic percentage. We conclude that the presence of contaminants exaggerates the O content on the diamond surface. In other words, rather than the O1s signal coming from the O-terminated carbon on the SCD, the O1s signal includes the contribution from the contaminants and the functionalized surface.

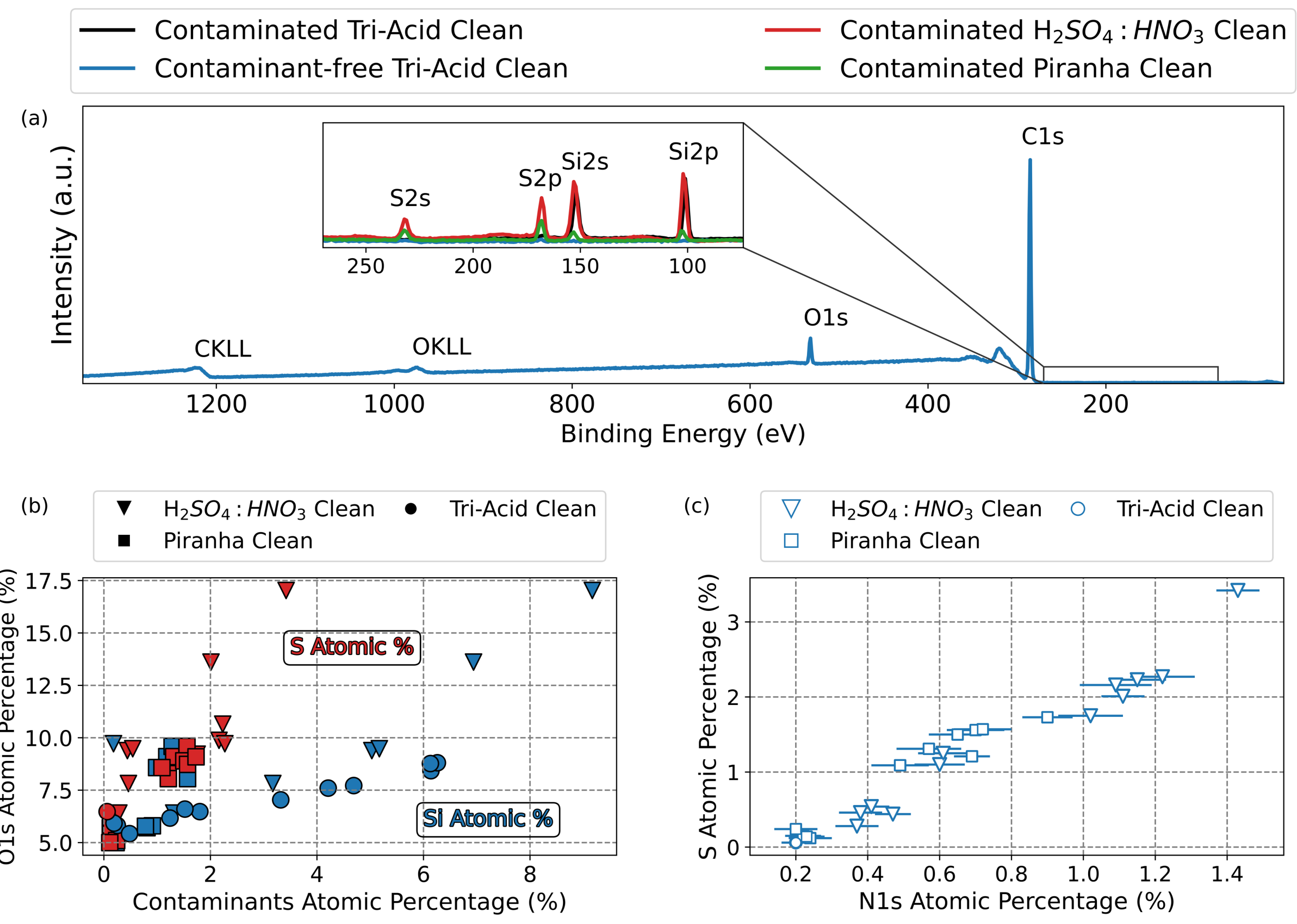

*Figure 2: Effect of molecular contamination on diamond surfaces: (a) Representative survey plot for clean and contaminated diamond for three different WCO methods. (b) O1s plotted against the most prominent contaminants identified in the study, silicon and sulfur. The results from each WCO method are labeled accordingly. (c) The sum of the S2s and S2p contributions plotted against N1s.*

The origin of the Si2s and Si2p signals is likely due to $SiO_2$, a contaminant that is ubiquitous in the everyday environment. Prior work by [30] and [31] suggested that sulfuric acid cleans lead to the formation of ammonium sulphate, as the residual amount of sulfuric acid on silicon surfaces readily reacts to nitrogen in the atmosphere. The ammonium sulphate then adheres to the diamond surface to which it is chemically bound and proves difficult to remove even with repeated water rinses [30]. Herein, we find evidence of the presence of these ammonium sulphate groups, shown on both Figure 2b and Figure 2c. In Figure 2b we observe how the data for the sulfur contribution is steeper than the silicon data. We conclude that this is due to the molecular origin of the sulfur being more O rich than the molecular form of the silicon signal. Furthermore, in Figure 2c the total sulfur atomic percentage (the sum of the S2s and S2p signals) is plotted against the N1s atomic percentage, which results in a positive correlation. This indicates that the presence of any sulfur on the diamond surface is usually accompanied by nitrogen. We find that the results shown on both Figure 2b and Figure 2c corroborate the identity of ammonium sulphate groups.

To mitigate the effects of the contaminants, we improved the purity of the acids used for the treatments (by switching from laboratory grade to optima grade acids), stored diamond samples in polypropylene-based, cleanroom-grade containers (avoiding the use of gelpaks) and performed a thorough clean of the diamond surfaces using multiple rinses with deionized water and optima-grade isopropyl alcohol prior to drying surfaces. As a result of these practices, all subsequent WCO results showed lower levels of contamination as compared to before. On average, the improved tri-acid clean has 0.10% of contamination observable on the diamond surface, while the nitric-sulfuric and piranha clean have 0.67% and 0.32% respectively. In turn, by mitigating the level of contamination from the tri-acid clean, it allowed us to garner a set of consistent oxidation data from all subsequent O-termination methods.

*3.2 XPS peak-fitting procedure to analyze functionalizations on the SCD surface*

Figure 3a displays the C1s spectra, as well the peak-fitting results for a tri-acid cleaned regular-polished diamond sample. CasaXPS was utilized to perform all peak fits with C1s narrow scan spectra. A universal Tougaard baseline was used as the background for all C1s peak fits. Voigt functions, in the form of a Gaussian-Lorentzian product were applied for all fits. Although several groups make mention of constraining the FWHM for all C1s peak deconvolutions to about 1 eV [19] [32] [33], we find here that the residual standard deviation (STD) is only minimized when the FWHM for the $sp^3$ peak is roughly 0.6 eV and constrained within 0.6 to 0.7 eV. FWHM values for all other chemical species were constrained to within 1.30 to 1.6 eV. We ascertain that the lower FWHM for the $sp^3$ C peak emanates from the bulk crystal structure and does not correspond to the shallowest layers of the diamond, which itself harbors only the $sp^2$ C and oxygen contributions [18]. Furthermore, it is known that the FWHM for the $sp^3$ C diamond component is smaller than the FWHM for the shallow topmost amorphous $sp^2$ C region [34].

Ideally, natural lineshapes emanating from XPS spectra should be Lorentzian, but Gaussian characteristics are introduced due to heterogeneity present in the sample [35] [36]. In this case, we find that the heterogenous nature of the diamond, i.e., a shallow amorphous $sp^2$ C component which is eventually followed by the bulk $sp^3$ C diamond [18], introduces lineshapes of mixed

Gaussian and Lorentzian character. These lineshapes take the form of a Gaussian – Lorentzian product function, labeled in Figure 3 as GL. Seeing as how the functional groups are confined to only the uppermost portions of the diamond, we imposed identical FWHM constraints and Gaussian quality on all the functional groups and $sp^2$ C, while giving the $sp^3$ C peak its own set of FWHM constraints and Gaussian character.

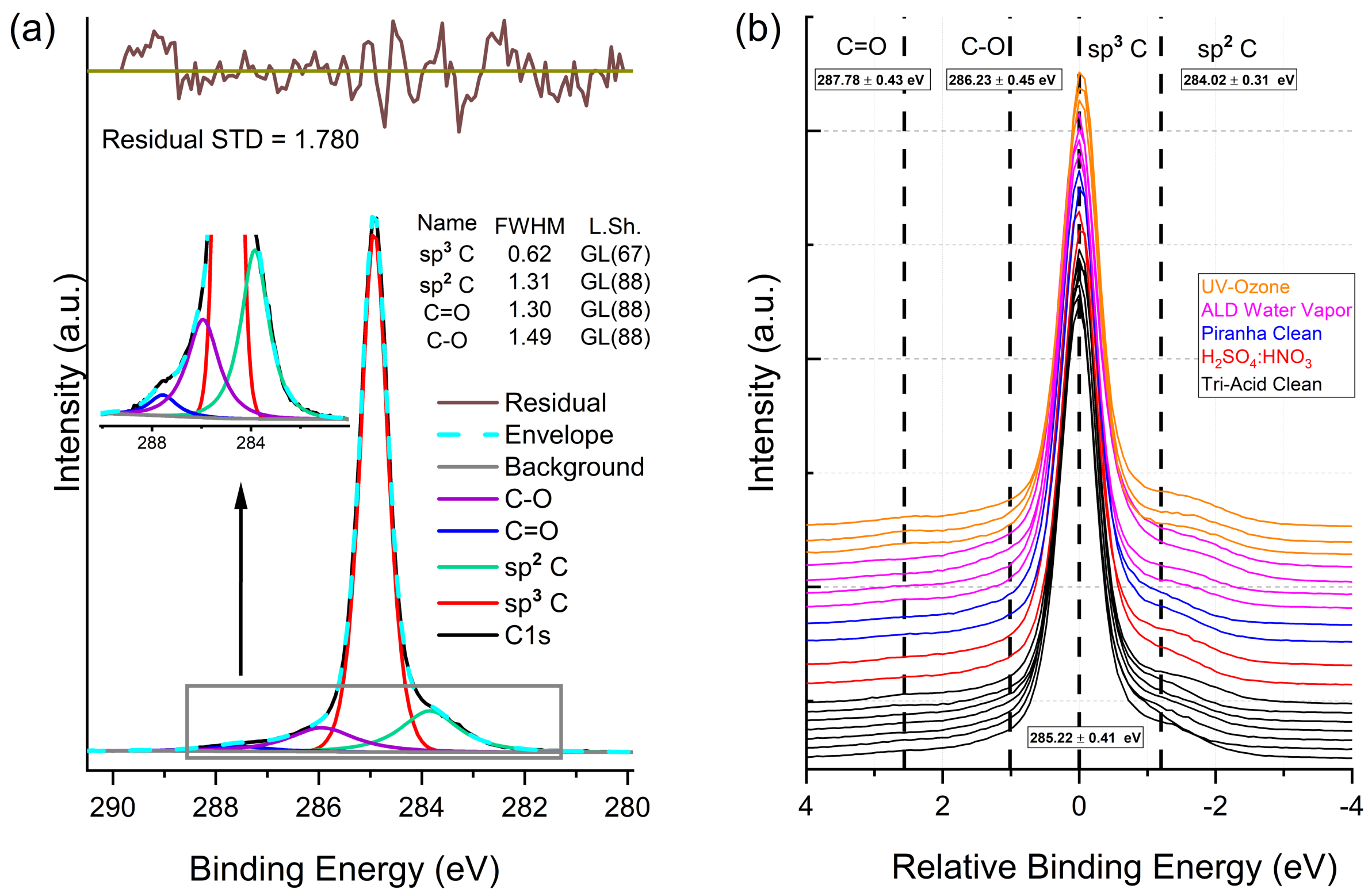

*Figure 3: (a) C1s peak fitting results from CasaXPS displaying the individual chemical contributions, the FWHM values, and the line shape for a tri-acid cleaned, regular-polished diamond sample. (b) Overlay plot of all C1s peaks, with their corresponding bond types and binding energy peak positions. The x-axis is plotted in terms of the relative binding energy, adjusted with respect to the bulk $sp^3$ C peak. Shown in the text boxes is the absolute value of all peak-positions.*

To generate chemically accurate C1s peak fits for all oxidized diamond data, approximate peak positions and identities for the four chemical species labeled in Figure 3a were taken from established literature values [17] [18] [37]. While a multitude of C1s plots were analyzed, Figure 3a, taken from a tri-acid cleaned sample, is representative of all SCD O-terminated peak fits analyzed in this study. The largest component among all oxidized SCD samples was the $sp^3$ C peak at a BE value of 284.93 eV. To the right of the $sp^3$ C peak, the $sp^2$ C peak was identified at 283.84 eV. A recent study has pointed to the existence of a native $sp^2$ C on SCD layer with a thickness of approximately 0.24 nm [18]. To the immediate left of the $sp^3$ C peak, at 285.94 eV, lies the singly bonded carbon peak, labeled as C-O in figure 3a. This peak consists of the contribution of all the singly bonded carbon groups, namely the ether and hydroxyl groups. Both computational and experimental work has supported the existence of O-terminated SCD surfaces comprised of at least two functional groups, namely the ether and ketone groups [38] [39]. Another computational study that has considered a completely hydrogen terminated surface and then incorporated ether, ketone, and hydroxyl groups found that, in order for a 100% ether or hydroxyl termination to be more energetically favorable than a complete H-termination, that ketone groups must consist of at least 17% of the surface coverage [40]. The peak at the far-left hand side, at 287.56 eV, is the doubly bonded C peak shown as C=O, representative of the ketone groups on the sample.

As the $sp^3$ C peak was the most prominent and bulk feature among all the oxygen-terminated samples, all other peaks were defined with respect to this bulk peak on the relative binding energy (RBE) scale, giving the $sp^3$ C peak a RBE of 0 eV as shown on Figure 3b. Figure 3b displays the raw C1s spectra of all oxidation treatments performed on the SCD samples, plotted in terms of the RBE. Shown in text is the average and standard deviation (STD) of the peak position in BE. The vertical dotted lines show the approximate RBE position of the chemical species that were deconvoluted from the C1s spectra.

Given the minimal amount of shifting observed for the bulk $sp^3$ C peak (less than 0.5 eV), we find that this is evidence that the flood gun proved successful in mitigating surface charging effects. The individual peak area contributions for all O-terminated diamonds for each of the chemical moieties can be found in the supplementary section of this text.

To emphasize, we note that to produce both physically meaningful and chemically accurate C1s peak fits representative of the functionalizations on the diamond surface, it is vital to first constrain the appropriate fitting parameters, such as the FWHM, peak positions, and lineshapes. Once these constraints are applied to the model, only then does it become appropriate to attempt to fit the data such that the goodness-of-fit metric, in this case the residual STD, reaches a global minimum [36]. For a more in-depth analysis involved in the process of C1s peak fitting the reader is referred to other texts [41] [42] [36] [43] [44].

*3.3 Relationship between O1s counts and the $sp^2$ content*

Figure 4 displays the relation between the $sp^2$ C peak area percentage and the O1s atomic percentage. All oxidation methods (with the exception of the HPC method) are shown in the figure, as well as their corresponding polishing treatments. All regular-polished WCO samples tend to cluster in the same region, which is indicative that these treatments produced roughly similar amounts of O1s and $sp^2$ C content. In contrast, the regular-polished DO procedures show lower amounts of $sp^2$ C, with O1s atomic percentages that are either equal to or greater than the WCO treatments. These results are consistent with prior work on PCD samples, as the $sp^2$ C content for PCD samples treated with WCO methods was shown to be higher than samples treated with DO methods [17].

With the regular-polished DO samples we also observe a correlation between the amount of $sp^2$ C and the O1s atomic percentage, before eventually plateauing out at the super-polished DO samples. Given how the super-polished samples all exhibit higher amounts of $sp^2$ C we note that the polishing process resulted in the formation of more $sp^2$ C on the shallowest parts of the diamond surface, an effect that has been studied in prior computational work [45]. Despite this, the O1s atomic percentage seems to not have increased and remains at a level comparable to the regular-polished UV/Ozone data.

Given these results, we conclude that the WCO treatments had high variance in regard to their $sp^2$ C content, but all of these treatments resulted in similar amounts of O1s atomic percentage. In comparison, the DO methods had lower amounts of $sp^2$ C and resulted in higher yields of O1s atomic percentage. However, the subsequent processing steps of super-polishing seems to have increased the $sp^2$ C content with only a modest increase in the O1s atomic percentage, which seems to have plateaued at a value roughly equal to the most effective regular-polished DO treatment. Prior work that has analyzed the layered structure of the SCD surface has shown that the most superficial portions of the diamond contain a rich O layer followed by amorphous $sp^2$ C [18]. The results in Figure 4 suggest that there is some packing efficiency related to how much O can fit in on the uppermost portion of the $sp^2$ C layer, as increasing the $sp^2$ C past 13% has not affected the O1s content.

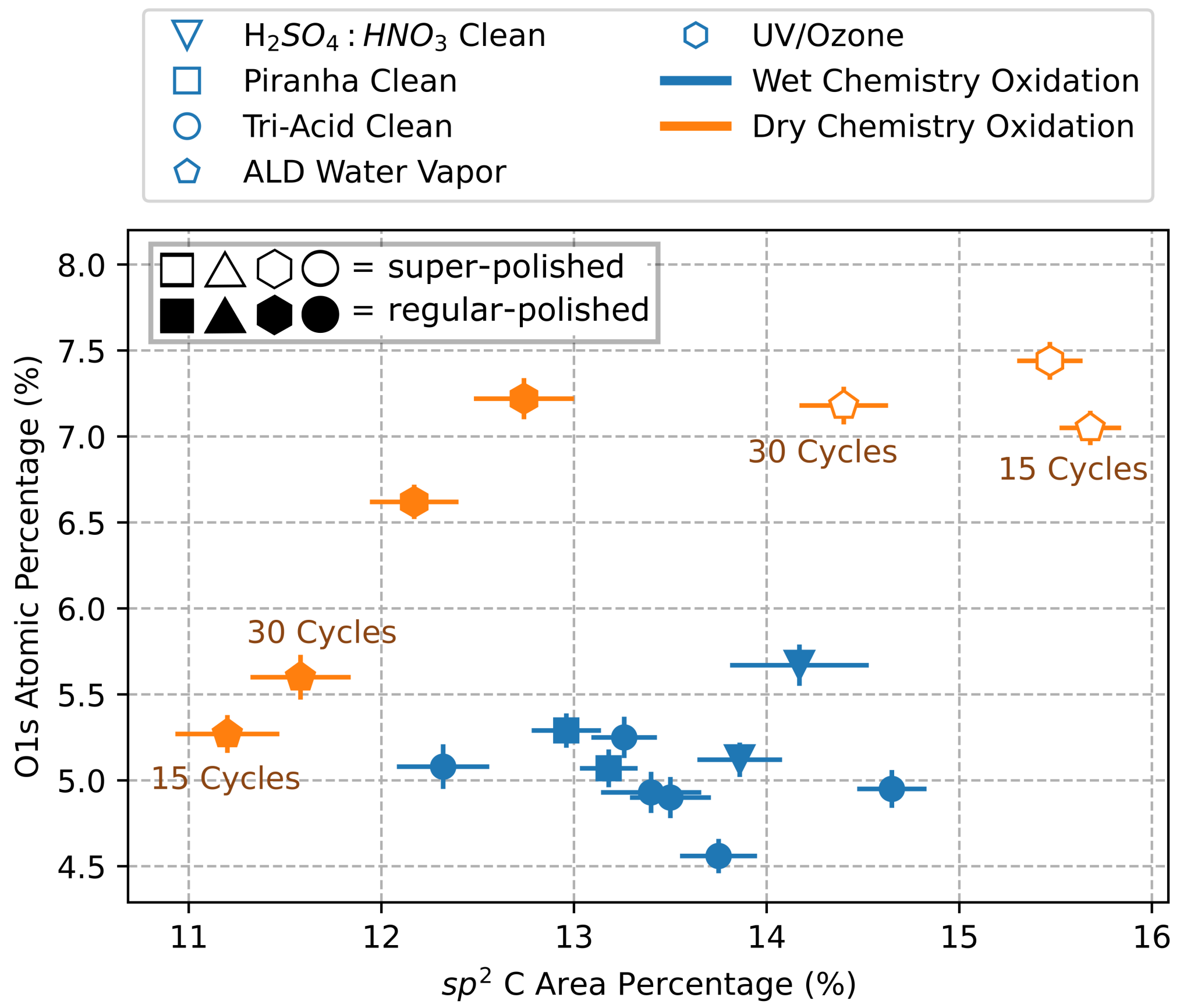

*Figure 4: $sp^2$ area plotted against O1s area measured for all wet chemistry and dry oxidation methods.*

The persistence of $sp^2$ C with all oxidized diamond samples is something that has been well documented in SCD samples [12], polycrystalline diamond (PCD) [17], and microcrystalline diamond (MCD) [46]. In Cobb et. al, it was shown that, despite applying a $H_2SO_4$:$KNO_3$ clean to MCD processed by laser micromachining, a layer of amorphous C remained on the surface, as verified through Scanning Transmission Electron Microscopy (STEM) [46]. Similarly, Li et. al showed that varying volumetric mixtures of $H_2SO_4$:$HNO_3$ cleans and a piranha clean led to visible $sp^2$ C content on PCD surfaces as verified through XPS C1s peak deconvolution, with the piranha clean and aqua regia solution yielding the most $sp^2$ C [17].

To ameliorate the persistence of $sp^2$ C content on the samples, we employed a H-termination technique via plasma treatment, as such treatments have been shown to produce chemically homogenous surfaces [47]. Afterwards, the HPC method was applied to the diamond sample. Figure 5 displays the C1s spectra as well as the C1s peak deconvolution results for the HPC sample. Peak identities and approximate positions for the chemical species identified from the HPC sample stem from Ghodbane et. al [48] [49] and Hoffman et. al [50]. The HPC data reveals the presence of $sp^3$ C, $sp^2$ C, C-H, C-O, and C=O bond types, with the $sp^3$ C and C-H bonds being the most predominant.

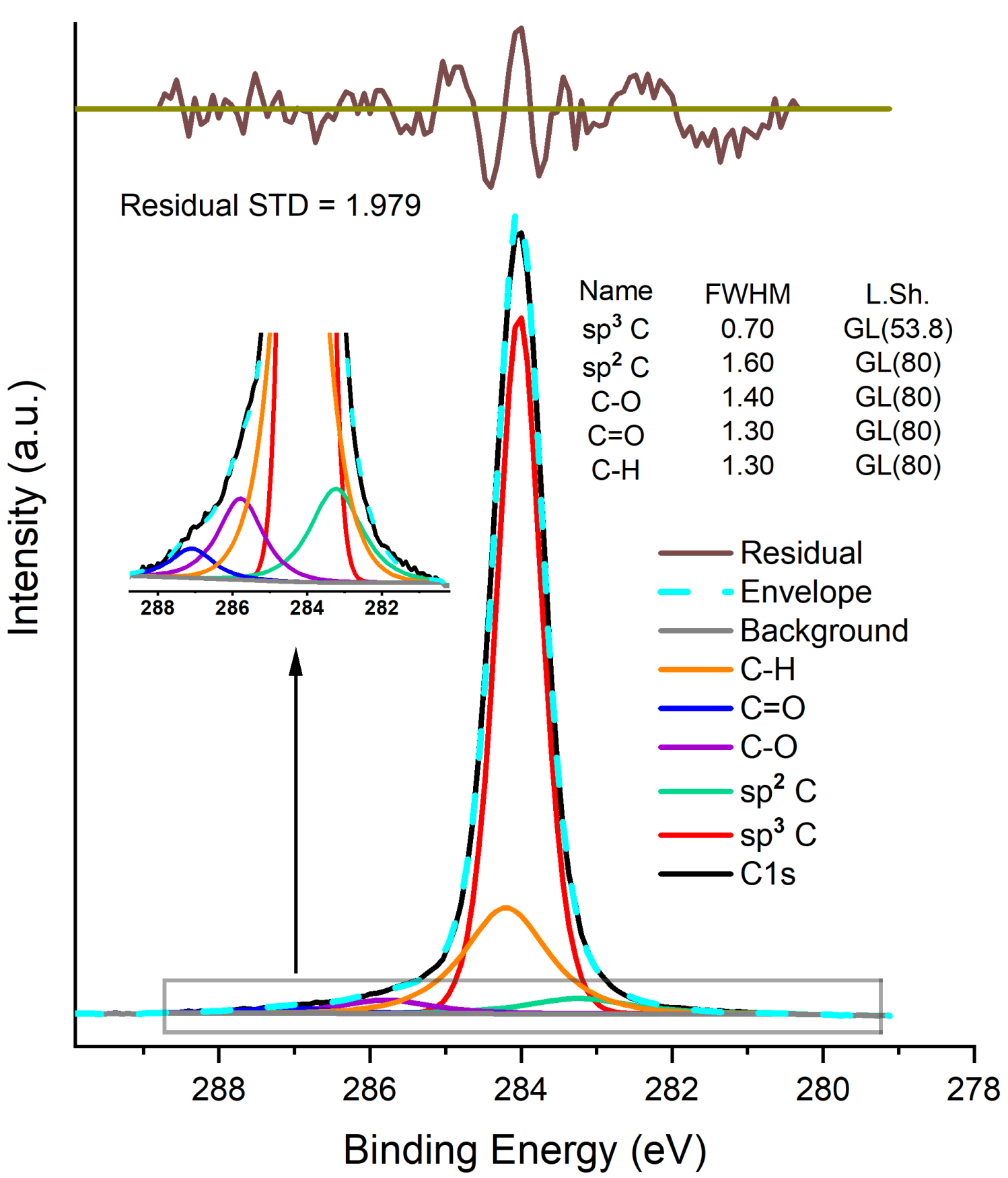

*Figure 5: (a) Peak-fitting results and information from hydrogen-termination plus carboxylation for the C1s peak.*

We also note the presence of peak shifting in the HPC sample which can be explained by the presence of C-H bonds on the diamonds surface. The H-bonds contribute to upward band bending which is caused by surface Fermi level pinning, which in turn reduces the barrier needed for electron emission from the $sp^3$ C peak [51] [52]. As a result, we see that the $sp^3$ C peak is lower in BE value, as the data in Figure 5 shows the $sp^3$ C peak position at 284.19 ± 0.16 eV.

The data from the HPC sample shows that treating the hydrogen-terminated sample with a subsequent acid-treatment was successful in producing the same functional groups which could be found on the non-H-terminated samples. However, the C-H peak remains dominant when compared to the other peaks and constitutes 21.39 ± 0.67% of the deconvoluted peak area. While the acid treatment has resulted in generating O functional groups, it was not successful in removing all the C-H bonds from the surface. This leads us to believe that the HPC method results in a partially functionalized surface, which is also evident by the decrease in O1s atomic percentage when compared to the non H-terminated samples, as those samples had O1s atomic percentage amounts of more than 4.84%, while the HPC sample had 3.47 ± 0.30%. Finally, we find that the lowest amount of $sp^2$ C peak area percentage, at 3.83 ± 0.50%, was found for the HPC method which we attribute to the etching of $sp^2$ C by hydrogen plasma during H-termination followed by the formation of C-H bonds on the diamond surface. For exact peak area percentages and counts for all chemical groups that were peak fitted from all C1s spectra in this study, the user is referred to the supplementary material of this text.

*3.4 Compilation of oxygen content from all surface treatments on single crystalline diamond*

Finally, the O1s atomic percentage results across ten treatment conditions are summarized in Figure 6. To put the O1s atomic percentage in context, using the NIST Effective-Electron-Attenuation Length Database [53], and assuming a terminal layer of $sp^3$ C on diamond, we determine that the surface C atoms on (100) diamond contributes to 6.4% of the C XPS counts for a pristine sample. Both ether and ketone groups can form with surface C atoms on the diamond (100) surface at a 1:1 O:C ratio. Therefore, if the XPS counts come only from O1s oxygen bound to the (100) diamond surface (in the form of surface ketone and ether groups) and C-C from $sp^3$ diamond, then a monolayer coverage of oxygen would correspond to the sum of C=O and C-O counts, or equivalently the O1s counts, being 6.0% of the total XPS signal. Although the idea of a monolayer can be useful for visualizing how oxygen can bond to a surface composed of $sp^3$ C, recent work suggests that oxidized diamond is multi-layered, with the most superficial layer consisting of most of the O content followed by the amorphous $sp^2$ C region, and then eventually transitioning into the $sp^3$ C diamond comprising the bulk of the material [18] [46]. Our observation of persistent sp2 C in the XPS signals from all our oxidized diamond samples is consistent with this model.

To better interpret the O1s atomic percentage values, we have included in Figure 6 the contaminant atomic percentage and the $sp^2$ C peak area counts, taken from the narrow scan C1s spectra peak-fitting results. The most right-hand side of Figure 6 also includes the number of samples used for each treatment which is denoted as "N". Final values for the O1s atomic percentages were calculated by averaging the survey values from all samples included within each processing method. The error bar for the O1s atomic percentage represents the error in the standard deviation from the mean and the uncertainty in the noise of the survey spectra. Likewise, on the left-hand side of the bar plot, $sp^2$ C area percentage values corresponding to N>1 have been calculated by averaging the $sp^2$ C values from all samples, with error bars calculated in a similar manner. The contaminant atomic percentage is taken to be the total sum of the individual atomic percentage of the Si2s, Si2p, S2s, S2p, and N1s contributions.

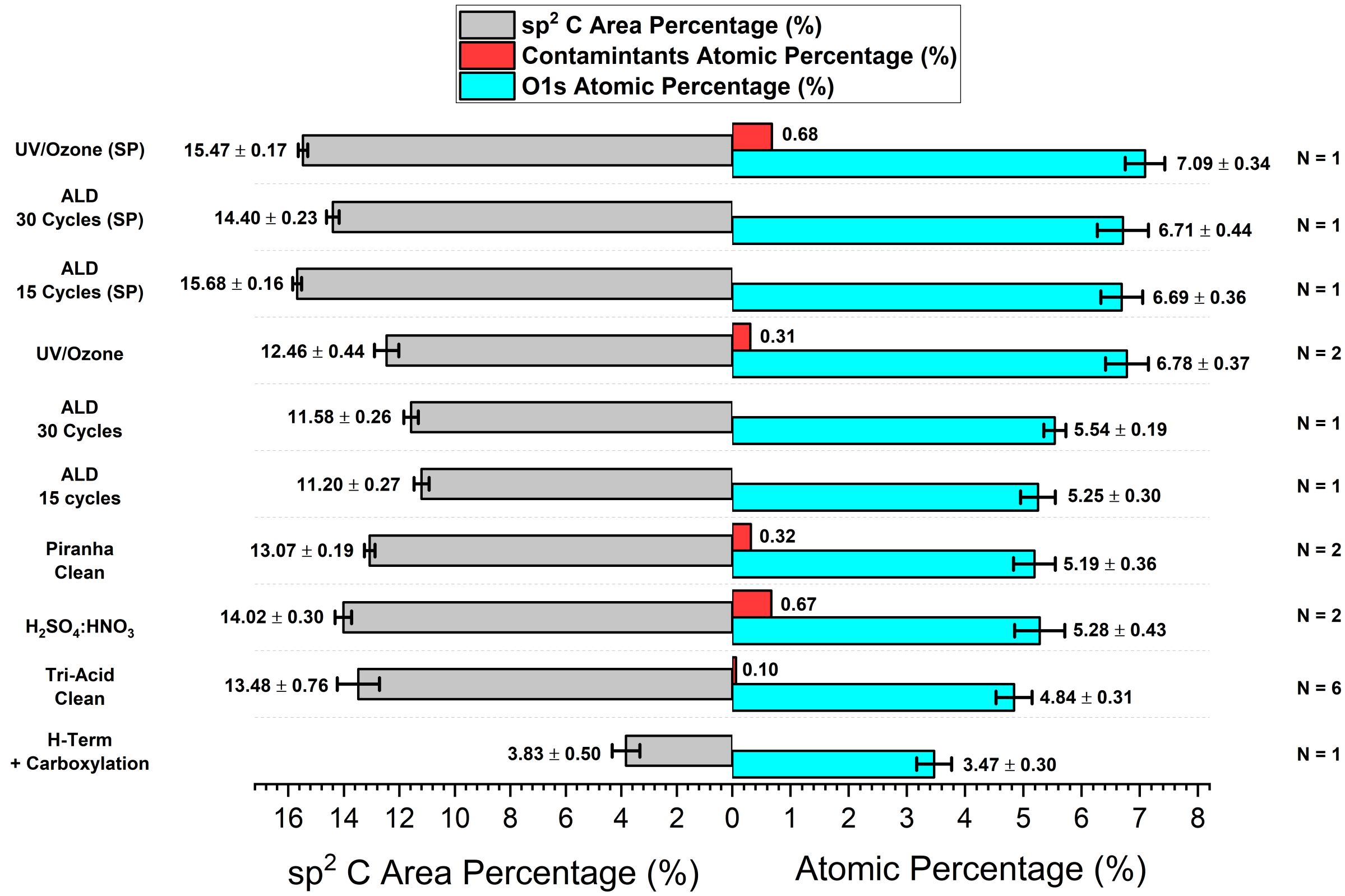

*Figure 6: Bar plot with all diamond treatment methods showing the O1s atomic percentage, Contaminant atomic percentage, and the sp2 C content that was present on the diamond samples from XPS measurement. The right-hand side with the value of N shows the number of samples that were used for each treatment.*

We identify the tri-acid clean as the cleanest WCO technique to O-terminate diamond, given that it has the lowest contaminant presence and the most repeatable O1s atomic percentage value, evidenced by the scale of its error bar when compared to the other methods. As elaborated upon in the prior section, the super-polished samples show elevated amounts of $sp^2$ C, when compared to the regular-polished samples, but the O1s content remains similar to the regular-polished UV/Ozone samples. We observed that further mechanical polishing of the diamond surface topologically homogenizes the diamond surface but tends to elevate the overall $sp^2$ C content [45]. This effect may be an important consideration in preparing oxidized surfaces in diamond quantum devices with shallow NV centers, as $sp^2$ C can be host to electron traps on the surface and when present in significant quantity has been shown to be detrimental to NV spin performance [54].

## 4. Conclusion

In summary, we have performed both WCO and DO treatments on (100) SCD and documented the effects of the $sp^2$ C content, the O1s atomic percentage, and the molecular contaminant level that resulted from the oxidation methods. In doing so, we have established a protocol of best practices that ensures sample preparation with low contamination from O-termination procedures. Furthermore, we have also put forth a comprehensive approach for reliable XPS peak-fitting of C1s spectra. While some of the most prominent methods for oxidizing diamond has been using wet chemistry (Table 1), the reported O atomic percentage values on the (100) SCD have a large variance, pointing to other factors that affect both the effectiveness of oxidation and the measured O content. Our work here provides insights on how the interpretation of XPS results and the methodology of sample preparation and XPS measurements could account for the variance in reported results.

By subjecting (100) SCD surfaces to several WCO and DO methods and using XPS measurements to analyze their surface content, we identified sulfur and silicon based molecular contaminants that exaggerate the O1s atomic percentage taken from survey data. Therefore, we encourage all future work of SCD oxidation to comment on the contaminant level of their samples. By comparing the $sp^2$ C and O1s content of all samples, we find that the super-polished samples have higher amounts of $sp^2$ C when compared to their regular-polished counterparts. Additionally, we observe that the O1s atomic percentage stays constant at over 7.0% with respect to an increase in $sp^2$ C.

Among the WCO methods used for this study, the tri-acid clean yielded the most consistent and reproducible oxygen amount, corresponding to the lowest spread in oxygen atomic percentage and the cleanest surface. DO methods, such as the photochemical oxidation via UV-illumination, and steam oxidation using ALD, are promising techniques that produce the highest oxygen atomic percentages measured in this study. Although the UV/Ozone method on both sets of regular-polished and super-polished samples produced relatively higher amounts of O1s atomic percentages when compared to the ALD water vapor method, we note that the UV/Ozone method was also prone to molecular contamination, with 0.68% and 0.31% for the super-polished and regular-polished samples respectively. Therefore, we conclude that the ALD water vapor method is the most promising oxidation technique overall, as this yielded the cleanest and highest amounts of O1s atomic percentages. Although oxygen and water vapor has been used prior to etch crystalline diamond surfaces, which in turn might damage the shallowest color centers within nanometers away from the surface, our ALD water vapor technique likely occurs at too low a temperature (250 ˚C) to cause any substantial damage to near-surface NV centers, as both oxygen gas and water vapor are only effective at etching diamond past 600˚C [55] [56] [57] [58]. Moreover, previous work on using oxygen gas at temperatures ranging from 445 - 450˚C to O terminate SCD (referenced as a DO method in Table 1) had shown that NVs as shallow as 4 nm had noticeably improved $T_2$ performance when compared to the tri-acid cleaned surface [12]. Finally, while the O1s atomic percentage for each WCO and DO method explored here is not highly sensitive to certain process parameters including the duration of the oxidation, the present work does not focus on process optimization which warrants future studies to be performed in combination with evaluating the effects on the quantum properties of near-surface color centers.

Acknowledgements

This work is supported by the U.S. Department of Energy, Office of Science, Basic Energy Sciences under Award #DE-SC0020313. Part of this work was performed under the auspices of the U.S. Department of Energy at Lawrence Livermore National Laboratory under Contract DE-AC52-07NA27344. Contributions by Robert J. Hamers, Cesar Saucedo, and Benjamin Bachman were supported by NSF CHE-1839174. The authors gratefully acknowledge use of facilities and instrumentation supported by NSF through the University of Wisconsin Materials Research Science and Engineering Center (DMR-1720415). RV and JTC gratefully acknowledge helpful discussions with Nathalie de Leon and Adrien Couet.

Data availability statement The data that support the findings of this study are available upon request from the authors.

Supplementary Information for "XPS analysis of molecular contamination and sp2 amorphous carbon on oxidized (100) diamond"

**S1:** The following section details the peak area percentages for each C1s fit and the binding energies for the functional groups performed on oxidized and H-terminated samples. Samples that have undergone further polishing prior to processing have been labeled as (SP).

*Table 1: Peak areas and counts for all identified functional groups of C1s peak deconvolution for all O-terminated diamond samples*

| Treatment | sp3 C Area (%) | sp3 C Area (Counts) | sp2 C Area (%) | sp2 C Area (Counts) | C-O Area (%) | C-O Area (Counts) | C=O Area (%) | C=O Area (Counts) |
|---|---|---|---|---|---|---|---|---|
| Tri-Acid Clean #1 | 74.02 ± 0.29 | 20375.88 ± 79.83 | 14.65 ± 0.21 | 4030.16 ± 57.77 | 8.46 ± 0.29 | 2328.03 ± 79.80 | 2.88 ± 0.21 | 791.93 ± 57.74 |
| Tri-Acid Clean #2 | 77.33 ± 0.28 | 21196.86 ± 76.75 | 13.26 ± 0.20 | 3635.45 ± 54.83 | 7.27 ± 0.30 | 1993.94 ± 82.28 | 2.14 ± 0.18 | 586.23 ± 49.31 |
| Tri-Acid Clean #3 | 75.44 ± 0.29 | 21638.30 ± 83.18 | 13.75 ± 0.17 | 3944.0 ± 48.76 | 9.12 ± 0.29 | 2617.40 ± 83.23 | 1.68 ± 0.29 | 483.20 ± 83.41 |
| Tri-Acid Clean #4 | 74.94 ± 0.32 | 21547.90 ± 92.01 | 13.40 ± 0.24 | 3851.86 ± 68.99 | 9.58 ± 0.25 | 2755.53 ± 71.91 | 2.08 ± 0.19 | 597.64 ± 54.59 |
| Tri-Acid Clean #5 | 74.66 ± 0.26 | 20328.81 ± 70.79 | 12.32 ± 0.31 | 3353.72 ± 84.39 | 11.14 ± 0.39 | 3034.21 ± 106.22 | 1.88 ± 0.19 | 511.10 ± 51.65 |
| Tri-Acid Clean #6 | 76.01 ± 0.28 | 19512.40 ± 71.88 | 13.50 ± 0.24 | 3466.35 ± 61.62 | 8.43 ± 0.27 | 2164.67 ± 69.33 | 2.06 ± 0.16 | 528.78 ± 41.07 |
| H2SO4:HNO3 #1 | 75.68 ± 0.24 | 21670.18 ± 68.72 | 13.86 ± 0.19 | 3968.72 ± 54.41 | 8.18 ± 0.34 | 2344.20 ± 97.44 | 2.27 ± 0.19 | 651.76 ± 54.55 |
| H2SO4:HNO3 #2 | 75.19 ± 0.30 | 21125.45 ± 84.29 | 14.17 ± 0.24 | 3980.10 ± 67.41 | 8.76 ± 0.39 | 2461.00 ± 109.57 | 1.89 ± 0.19 | 530.66 ± 53.35 |
| Piranha Clean #1 | 76.01 ± 0.24 | 22336.89 ± 70.53 | 13.18 ± 0.22 | 3870.95 ± 64.61 | 8.98 ± 0.32 | 2638.98 ± 94.04 | 1.84 ± 0.14 | 540.83 ± 41.15 |
| Piranha Clean #2 | 74.05 ± 0.31 | 21422.25 ± 89.68 | 12.96 ± 0.17 | 3747.62 ± 49.16 | 10.95 ± 0.37 | 3168.50 ± 107.06 | 2.04 ± 0.19 | 590.91 ± 55.04 |
| ALD Water Vapor 15 Cycles | 76.84 ± 0.31 | 22852.99 ± 92.20 | 11.20 ± 0.22 | 3329.59 ± 65.40 | 9.89 ± 0.27 | 2942.52 ± 80.33 | 2.07 ± 0.16 | 614.72 ± 47.51 |

| ALD Water Vapor 30 Cycles | 76.36 ± 0.32 | 20615.81 ± 86.39 | 11.58 ± 0.28 | 3125.07 ± 75.56 | 9.72 ± 0.36 | 2623.35 ± 97.16 | 2.35 ± 0.25 | 634.75 ± 67.53 |
|---|---|---|---|---|---|---|---|---|
| UV/Ozone #1 | 74.44 ± 0.36 | 19582.24 ± 94.70 | 12.74 ± 0.30 | 3350.17 ± 78.89 | 9.77 ± 0.35 | 2570.03 ± 92.07 | 3.05 ± 0.18 | 803.36 ± 47.41 |
| UV/Ozone #2 | 75.59 ± 0.35 | 21545.03 ± 99.76 | 12.17 ± 0.25 | 3467.77 ± 71.24 | 9.38 ± 0.29 | 2673.43 ± 82.65 | 2.86 ± 0.19 | 814.94 ± 54.14 |
| ALD Water Vapor 15 Cycles (SP) | 75.44 ± 0.29 | 30923.71 ± 118.87 | 15.68 ± 0.19 | 6426.02 ± 77.87 | 6.74 ± 0.16 | 2762.18 ± 65.57 | 2.14 ± 0.18 | 877.73 ± 73.83 |
| ALD Water Vapor 30 Cycles (SP) | 77.35 ± 0.29 | 30303.69 ± 113.61 | 14.40 ± 0.22 | 5642.02 ± 86.20 | 6.13 ± 0.18 | 2403.24 ± 70.57 | 2.11 ± 0.19 | 827.22 ± 74.49 |
| UV/Ozone (SP) | 74.32 ± 0.25 | 29510.70 ± 99.27 | 15.47 ± 0.19 | 6140.36 ± 75.41 | 6.54 ± 0.24 | 2595.44 ± 95.24 | 3.68 ± 0.17 | 1460.26 ± 67.46 |

*Table 2: Binding energy peak positions for all identified functional groups for all O-terminated diamond samples*

| Chemical Bond | Peak Position Average (eV) | Peak Position Error (eV) |
|---|---|---|
| sp3 C | 285.22 | 0.41 |
| sp2 C | 284.02 | 0.31 |
| C=O | 287.78 | 0.43 |
| C-O | 286.23 | 0.45 |

*Table 3: Peak area information for all C1s peak deconvolution for H-terminated plus oxidation treated diamond samples, based on the procedure from Wang et al [1]*

| Bonding Type | HC |
|---|---|
| Sp3 Area (%) | 70.91 ± 0.98 |
| Sp3 Area (counts) | 19257.12 ± 266.14 |
| Sp2 Area (%) | 3.83 ± 0.48 |
| Sp2 Area (counts) | 1038.84 ± 130.19 |
| C-H Area (%) | 21.39 ± 0.67 |
| C-H Area (counts) | 5808.39 ± 181.94 |

Supplementary

| C-O Area (%) | 2.90 ± 0.32 |
| --- | --- |
| C-O Area (counts) | 788.52 ± 87.01 |
| C=O Area (%) | 0.98 ± 0.16 |
| C=O Area (counts) | 265.85 ± 43.40 |

*Table 4: Binding energy peak positions for all identified functional groups from H-terminated plus oxidized diamond samples*

| Chemical Bond | Peak Position (eV) |
| --- | --- |
| sp3 C | 284.03 |
| sp2 C | 283.23 |
| C-H | 284.20 |
| C-O | 285.78 |
| C=O | 287.09 |

**S2:** Shown below are the AFM images for the regular-polished ALD water vapor treated diamond samples at 15 and 30 cycles.

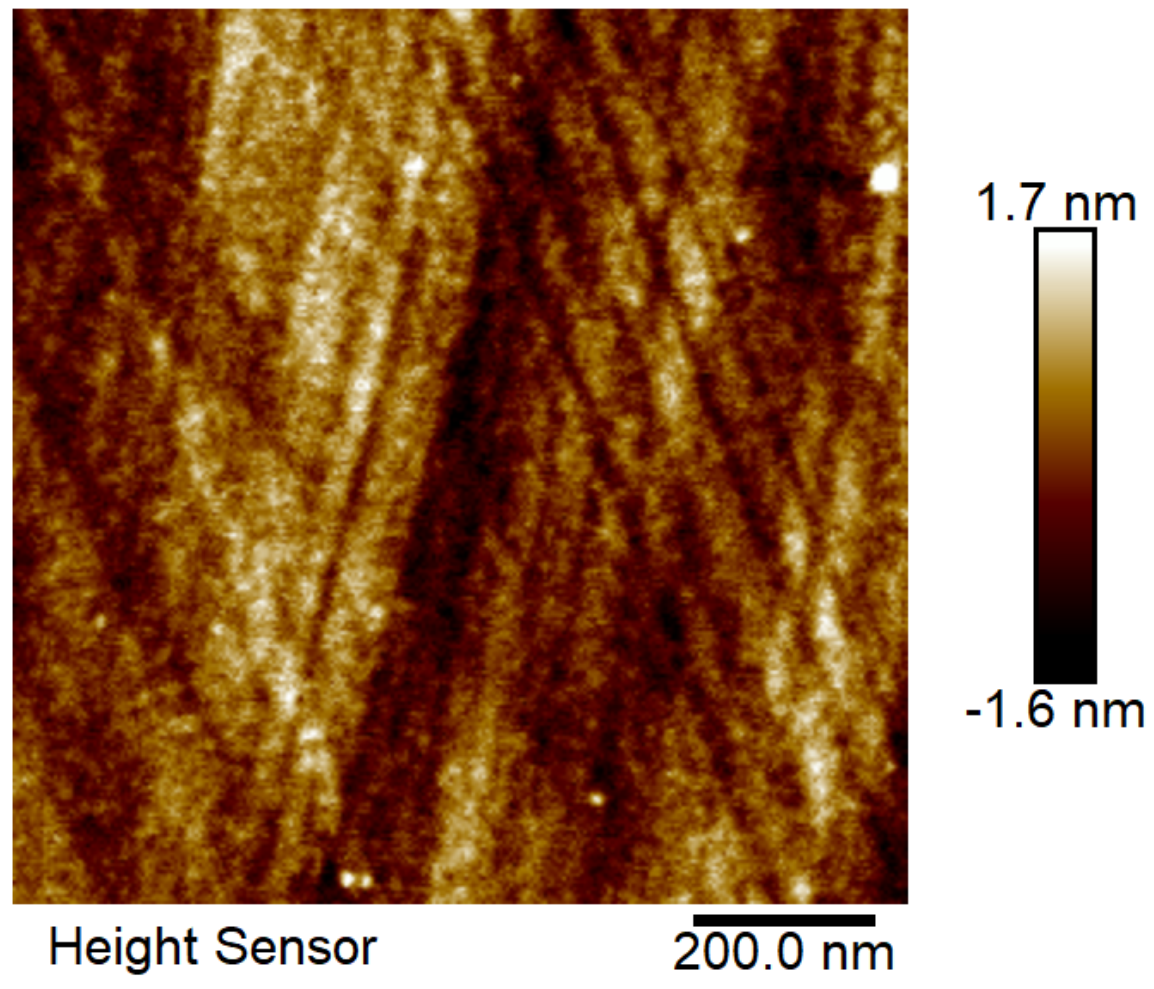

*Figure 1: AFM image of the regular-polished ALD water vapor sample with 15 cycles.*

Supplementary

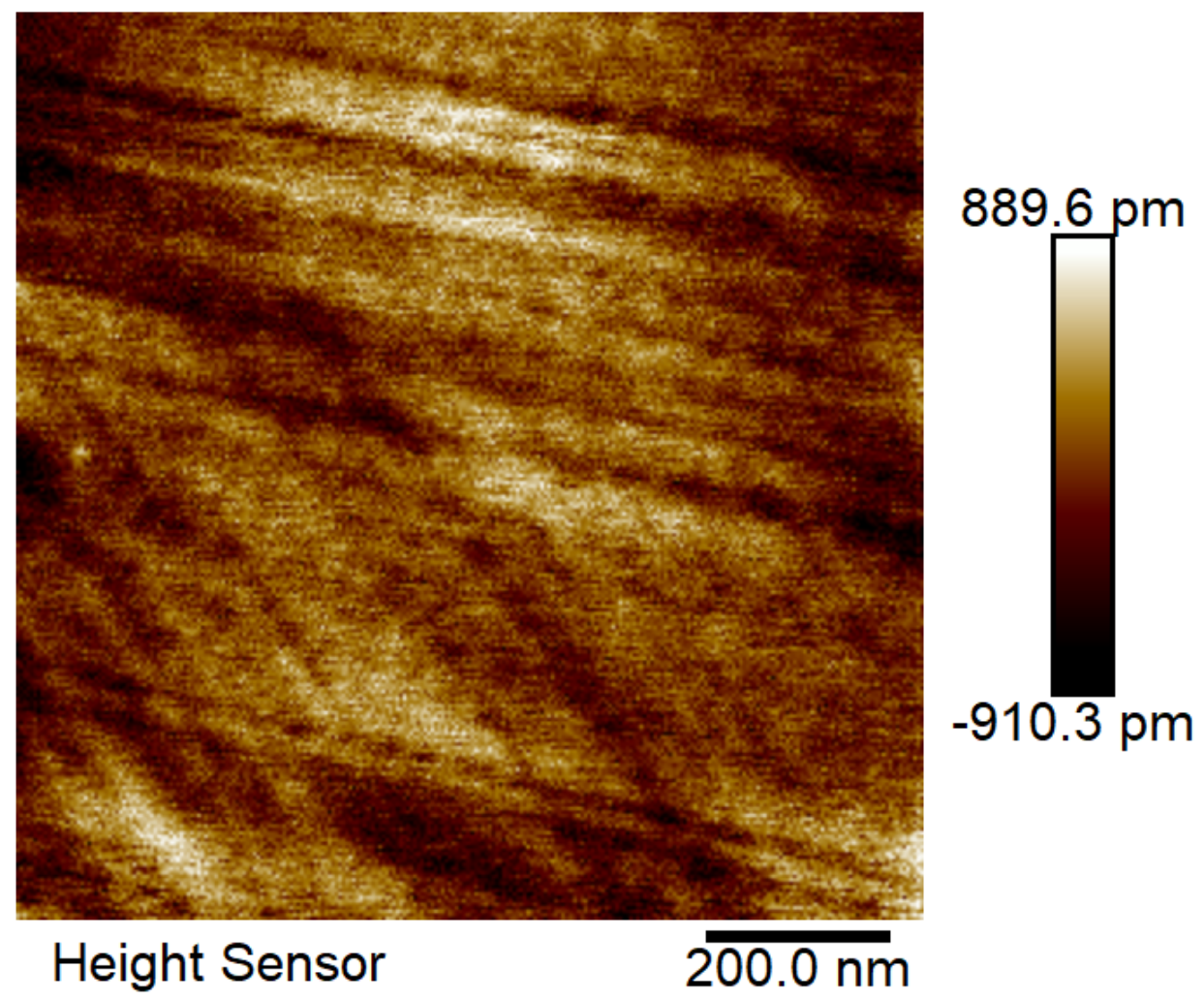

*Figure 2: AFM image of the regular-polished ALD water vapor sample with 30 cycles.*

References for the Supplementary Section